\documentstyle[pre,aps]{revtex}
\input psfig
\def\nv{{\bf n}}
\def\xv{{\bf x}}
\def\gradv{{\mbox{\boldmath{$\nabla$}}}}
\def\Gv{{\bf G}}
\def\uv{{\bf u}}
\def\Nv{{\bf N}}
\def\qv{{\bf q}}
\def\vv{{\bf v}}
\def\Angstrom{{\AA}}
\begin{document}
\twocolumn[\hsize\textwidth\columnwidth\hsize\csname@twocolumnfalse%
\endcsname
\title{Soft Condensed Matter Physics}
\author{T.C.~Lubensky}
\address{Department of Physics and Astronomy, University of Pennsylvania,
Philadelphia, PA 19104}
\draft
\date{\today}
\maketitle
\begin{abstract}
Soft condensed matter physics is the study of materials, such as fluids,
liquid crystals, polymers, colloids, and emulsions, that are ``soft" to
the touch.  This article will review some properties, such as the dominance
of entropy, that are unique to soft materials and some properties such as
the interplay between broken-symmetry, dynamic mode structure, and
topological defects that are common to all condensed matter
systems but which are most easily studied in soft systems.
\end{abstract}
\vskip2pc]
\section{Introduction}
In recent years soft condensed matter physics, or simply soft physics,
has emerged as an
identifiable subfield of the broader field of condensed matter physics.
As its title implies, it is the study of matter that is ``soft," i.e., of
materials that will not hurt your hand if you hit them.  This is in
contrast to ``hard" materials such as aluminum or sodium chloride that are
generally associated with the field of solid state physics.  Though the
term soft physics has only recently gained acceptance, its purview
is vast. It subsumes all of fluid physics, including both microscopic
structure and macroscopic phenomena such as hydrodynamic flow and
instabilities.  It includes liquid crystals and related materials with
their vast variety of broken-symmetry states.  It includes colloids,
emulsions, microemulsions, membranes, and a large fraction of biomaterials.
It is a field that presents fundamental scientific challenges and one that
has substantial economic impact.  In this article, I will give a brief
overview of some fundamental problems of soft condensed
matter physics\cite{ChaikinL}.
\par
The defining property of soft materials is the ease with which they respond
to external forces.  This means not only that they distort and flow in response
to modest shears but also that thermal fluctuations play an important if not
dominant role in determining their properties.  They cannot be described
simply in terms of harmonic excitations about a quantum ground state as most
hard materials can.  There are soft materials that possess
virtually every possible symmetry group, including three-dimensional
crystalline symmetries normally associated with hard materials and many others
not found at all in hard materials.  Ordered phases of soft materials can
easily be distorted, making it possible to study and to control states far
from equilibrium or riddled with defects. Thus, soft materials offer an ideal
testing ground for fundamental concepts, involving the connection between
symmetry, low-energy excitations, and topological defects, that are at the
very heart of physics.
\par
In this article, I will discuss four broad problems that reflect the
richness of soft physics: entropic forces and entropically
induced order, broken symmetries, topological defects, and membranes.
This is by no
means an exhaustive list; it does not, for example, include nonequilibrium and
non-linear phenomena or the vast field of polymer physics; it is, however, a
list that has general applicability to hard as well as soft physics.
\section{The Triumph of Entropy}
\subsection{Introduction}
Phases in thermodynamic equilibrium correspond to minima of the free energy
$F = E - TS$, where $E$ is the internal energy, $T$ is the temperature, and
$S$ is the entropy.  In hard materials, $E$ tends to dominate over entropy:
to a good approximation, internal energy determines the structure of
equilibrium phases, and thermal fluctuations can be treated as perturbations
about a minimum-energy phase.  In soft systems, quite the opposite may be
true: internal energy may either be small compared to $TS$, or it may not
depend at all on configurational changes of the system.  In the latter case,
equilibrium states are those that maximize the entropy rather than minimize
the internal energy. In addition, deviations of the entropy from its
equilibrium maximum value create forces whose effects are every bit as real as
those arising from the gradient of a potential.  Perhaps the most familiar of
such entropic forces is that required to stretch a polymer\cite{depol}.
A polymer can be
modeled as a sequence of $N$ freely joined segments of length $l$.  The
entropy of such a chain is a maximum if it is completely unconstrained.
Constraining its ends to have a separation $R$ leads to an entropy
reduction of $\Delta S = - 3 R^2/(2 Nl^2)$,
a free energy increase of $\Delta F = 3 T R^2/(2Nl^2)$, and a force
$f = - \partial F/\partial R = - 3 TR/(2N l^2)$.  In
this section, we will explore some systems where entropy determines structure
and interparticle forces.
\subsection{Hard Spheres}
\par
The interaction potential between spherical atoms such as the noble gases
consists of a long-range Van der Waals attractive part and a short-range
repulsive part arising largely from the Pauli principle.  Though the
formation of crystals at low-temperature depends critically on the
existence of the attractive part of the potential, the properties of the
liquid phases are determined to a large degree by the repulsive part,
which is well modeled by a hard-sphere interaction that is infinite for
interparticle separations less the particles' diameter and zero otherwise.
The hard-sphere gas was originally introduced as a mathematically simple
model to describe fluid phases.  Colloidal dispersions\cite{colloids} of
polystyrene spheres with radii ranging from $0.07\mu$m to $4\mu$m now provide
nearly prefect experimental realizations of hard-sphere
models.  They provide marvelous
laboratories in which to test a variety of theoretical predictions.  They have
also provided  unexpected results.
\par
In a hard-sphere system, the internal energy is zero in every allowed
configuration. Interparticle forces and the free energy are determined
entirely by entropy, which depends on the fraction $\phi$ of the total volume
occupied by the hard spheres.  At low volume fraction, collisions between
particles are rare, and the system is an ideal gas.  As volume fraction
increases, particle motion is increasingly restricted by collisions with
neighbors.  At close-packing densities, all particle motion is arrested.
A remarkable feature of hard-sphere systems is that there are two
close-packing densities: the hexagonal-close-packing density, $\phi_h =
0.7405$, and the random close-packing density, $\phi_c = 0.638$.  At
random close-packing, there is a random arrangement of particles such that
every particle has contacts with other particles that prevent its motion.  It
is important that $\phi_c < \phi_h$.  Imagine expanding a
hexagonal-close-packed lattice so that
its lattice structure is maintained.  Under this
expansion,
$\phi$ will decrease, and each particle will move freely in a cage centered
at a lattice site of the expanded lattice.  Because the particles are free to
move, they have a nonzero entropy.  Thus, such an expanded lattice at $\phi =
\phi_c$ will clearly have a higher entropy than the random-close packed
structure of the same volume fraction\cite{glass}.
This implies that as volume fraction is
increased from the liquid phase, it becomes entropically favorable for the
system to form a periodic crystal rather than a random fluid structure: there
is an entropically driven first-order liquid-to-solid
transition\cite{comcoex,expcoex}.   On the phase boundary, a liquid with
volume fraction $0.495$ coexists with an FCC crystal with volume fraction
$0.545$.
\par
Colloidal crystals have extremely small shear moduli compared to hard
solids.  Since the only energy in the problem is the temperature $T$,
dimensional analysis dictates that the shear modulus $\mu$ must be of
order $k_B T/v_0$ where $v_0$ is the free volume per particle, which is of
order $(\phi - \phi_h) a^3 \approx 0.1 a^3$ where $a$ is the lattice
spacing.  For $a$ somewhat less than a micron, this yields $\mu \sim
1 $ dyne/cm$^2$, $10^{12}$ times smaller than for a hard solid like
aluminum!.  In charged colloids, Coulomb interactions will increase $\mu$.
The small value of $\mu$ makes it possible to study phenomena like shear
melting of a crystal\cite{melting}.
\par
\begin{figure}
\centerline{\psfig{figure=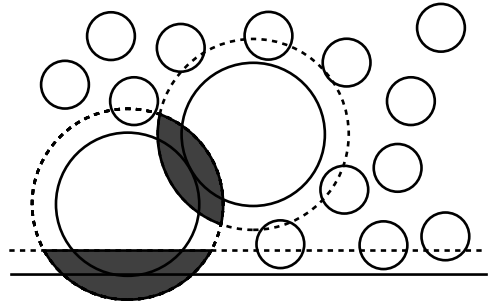}}
\caption{Drawing of two large spheres, small spheres, and a wall.  The full
lines indicate the hard-sphere radii of the spheres.  The dotted lines
indicate the volume around an {\it isolated} large sphere or the isolated
wall that centers of small spheres cannot enter.  The shaded regions
indicate the additional volume available to small spheres when two large
spheres touch or when a large spheres touches a wall relative to the volume
available when large spheres are far from each other and from any walls.}
\label{depletion1}
\end{figure}

Mixtures of hard spheres of different sizes provide compelling examples of
the power of entropy.  Consider first mixtures of large spheres  with radius
$a_L \approx 4\mu$m and small spheres with radius $a_S \approx 0.1\mu$m.  At
comparable volume fractions, there will be far more small spheres than large
spheres, and the entropy of the small spheres will dominate.  Thus to a good
approximation, the large spheres will adopt those configurations that maximize
the entropy of the small spheres.  The volume excluded to small spheres in
the presence of two large spheres is smaller if the two spheres touch than
if they are far apart as shown in Fig.\ \ref{depletion1}.  Thus, there is
an entropically induced attractive force between the large
spheres\cite{depl}.  This is called a {\it depletion} force.  It is of
considerable importance in colloids and emulsions and in biological
systems\cite{bio}.
The particles involved do not have to be hard spheres.  The small particle
could, for example, be a polymer\cite{pol} or a micelle.
\par
There is more free volume available to small particles if a large
particle is in contact with the surfaces of a container than if it is
removed from that surface. There is thus an entropic attraction of large
particles to surfaces.  Experiments\cite{surface} show not only that large
particles segregate at surfaces but that they do so at sufficient density to
produce a crystalline phase.  In the jargon of modern statistical physics,
entropic forces cause a crystalline large-particle phase to wet a fixed
surface.
\par
The above discussion, based on a pair-wise attractive interaction between
large spheres would lead one to expect that large and small particle would
phase separate with increasing volume fraction.  The experimental situation
is more complicated.  Under appropriate conditions, equilibrium crystalline
alloys\cite{alloy,theory} of large and small crystals, sometimes with very
large unit cells (e.g. $AB_{13}$), can form.  The entropic mechanism for this
effect is clear.  If the large spheres form a one-component, nearly
close-packed lattice, all of the volume between large spheres is denied to the
large spheres.  A slight expansion of the large-sphere lattice to allow entry
of small spheres into its open spaces can enhance the small-sphere entropy.
\subsection{Hard rods}
\par
\begin{figure}
\centerline{\psfig{figure=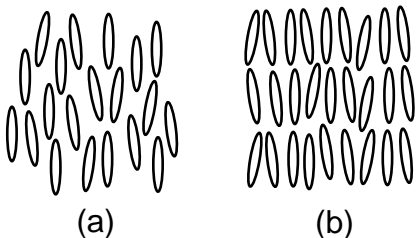}}
\caption{Schematic representation of (a) the nematic liquid crystalline phase
in which rod-like molecules (nematogens) align along a common direction while
their centers of mass diffuse freely as in a fluid and (b) the smectic-$A$
phase in which molecular centers of mass segregate into a one-dimensional
stack of two-dimensional fluid layers with molecular axes aligned on average
along the layer normals.}
\label{lcfigs}
\end{figure}

In the nematic liquid-crystalline phase\cite{LC}, the long axes of
bar-like molecules (Fig.\ \ref{lcfigs}a) called {\it nematogens} align on
average along a common direction specified by a unit vector $\nv $ called the
director.  Their centers of mass are, however, randomly distributed as in an
isotropic fluid. As first pointed out by Onsager\cite{Onsager}, the existence
and stability of this phase is due in large part to entropy. A hard rod has
both translational and rotational entropy.  Motion of randomly positioned and
oriented match sticks in a box of fixed volume  is strongly inhibited even at
a relatively small volume fraction.  The same is true for molecules.  At
high volume fractions, far more translational motion is possible if all of the
molecules align along a common direction.  The formation of the aligned
nematic phase results from an increase in translational entropy
that exceeds the
reduction of rotational entropy produced by molecular alignment.
Using a variational approach, Onsager calculated that isotropic and nematic
phases of solutions of spherocylinders of length $L$ and diameter $D$
coexist at respective volume fractions $\phi_i = 3.3 D/L$ and $\phi_n = 4.2
D/L$.
\par
Our experience with matches (or perhaps the childhood game ``pick-up-sticks")
makes the Onsager explanation for the existence of the nematic phase
extremely plausible.  What is more surprising is that hard-rod entropy also
leads to the smectic phase (Fig.\ \ref{lcfigs}b) in which the molecular centers
of mass segregate into a one-dimensional stack of two-dimensional fluid
layers. This result, predicted by computer simulations\cite{smsimul}, has been
verified by experiments\cite{TMV} on solutions of tobacco mosaic viruses, which
are almost ideal rigid rods.  Small spherical particles added to a solution of
hard rods induce an attractive depletion force between the rods.  However, as
in the case of mixtures of two different size spheres, mixtures of rods and
spheres can lead to periodic structures with complex multi-particle unit
cells\cite{Seth}.
\section{Broken Symmetries}
\subsection{Nematic Liquid Crystals: A Tutorial Example}
\par
Nematogens can be viewed as long rods.  At high temperature or low density,
these molecules rotate freely about all axes and form a homogeneous,
isotropic fluid.  In the nematic phase, the molecules align along a uniform
director, which can point in an arbitrary direction.  The isotropic phase is
invariant under arbitrary translations and rotations, i.e., under all of the
operations of the Euclidean group.  The nematic phase picks out a particular
direction for molecular alignment.  It is still invariant under arbitrary
translations, but it is invariant only under rotations about the axis
parallel to $\nv$.  the nematic phase has lower symmetry than the isotropic
phase: it is a {\it broken-symmetry} phase.
\par
This reduction in symmetry can be quantified by an {\it order parameter}
that is nonzero in the nematic phase and zero in the isotropic phase.  A
convenient order parameter, $Q_{ij}$, for the nematic phase can be defined as
follows.  Let $\nu_i$ be the unit vector specifying the direction of the
long axes of the nematogens.  Then $Q_{ij} = \langle \nu_i \nu_j - {1 \over
3}\delta_{ij} \rangle = S (n_i n_i - {1 \over 3} \delta_{ij})$ where
$\langle \,\,\rangle$ denotes an equilibrium ensemble average and $S = {1
\over 2}\langle 3 (\nu \cdot \nv )^2 - 1 \rangle$. This order parameter
reflects a broken rotational symmetry, yet like the nematic phase, it is
invariant under the inversion operation, $\nv \rightarrow - \nv$.
Different directions of $\nv$ define different ordered equilibrium
states.  Transformations between these states are produced by rotations
(through some angle $\theta$), which are symmetry operations of the
isotropic phase.  The order parameter $Q_{ij}$ transforms under the $l=2$
irreducible representation of the rotation group.
\par
Since there
is no energetically favored direction for
$\nv$, uniform rotations of a nematic will not change its free energy $F$.
Spatially nonuniform distortions will, however, increase $F$.  Since the
energy of these distortions must go to zero with wavenumber $\qv$, $F$ can be
expanded in a power series in gradients of $\nv$.  The result is the Frank
free energy
\begin{eqnarray}
F_n & = & { 1 \over 2} \int d^3 x \{K_1 (\gradv \cdot \nv )^2 + K_2 [\nv \cdot
(\gradv \times \nv ) ]^2 \nonumber \\
& & \qquad + K_3 [\nv \times (\gradv \times \nv )]^2 \} \nonumber \\
& \approx & {1 \over 2} K \sum_\qv q^2 |\delta \nv ( \qv ) |^2 ,
\label{frank}
\end{eqnarray}
where $K_1$, $K_2$, and $K_3$ are rigidity moduli or elastic constants
associated, respectively, with splay, twist, and bend distortions of the
nematic.  The elastic constants have
units of force and have a
typical magnitude of order $5 \times 10^{-7}$ dynes.  The
second form of $F$ is the harmonic limit with $\nv \approx (\delta n_x, \delta
n_y , 1)$ and equal elastic constants ($K_1 = K_2 =K_3 = K$). This free
energy implies a number of important properties of the nematic state.
First, there are low-energy distortions of the nematic phase with energies
$\epsilon_q = K q^2 $.  The dependence of this energy on $q^2$ is a direct
consequence of of the fact that the nematic phase breaks a continuous
(rotational) symmetry.  Second, the equipartition theorem states that
fluctuations in $\nv $ satisfy $\langle |\delta \nv ( \qv ) |^2 \rangle =
k_BT/K q^2$.  These fluctuations reduce order and tend to restore rotational
isotropy.  If $K$ were zero, fluctuations in $\nv$ would diverge and destroy
nematic order.  Thus, the existence of the broken-symmetry nematic state
requires the existence of a nonzero rigidity $K$.  Finally, the nematic phase
can transmit torques $\tau_i = \delta F/\delta n_ i = - K \nabla^2 \delta n_i$
even though it cannot support shear stresses as a solid can.
\par
The broken continuous symmetry leads to the existence of new low-frequency
dynamical modes\cite{Forster,MPP} in addition to low-energy static
excitations.  This is the Goldstone theorem\cite{Goldstone}.  An isotropic
fluid has five hydrodynamic modes (positive and negative frequency
longitudinal sound, two velocity diffusion, and one energy diffusion mode)
associated with the five conservations laws for mass, energy, and momentum.  A
nematic has two additional director modes (which mix with velocity diffusion
modes) associated with the two independent directions that a nematic can be
rotated. There is a frictional resistance, characterized by a friction
coefficient $\gamma \approx 0.1$ poise with units of viscosity, to the
rotation of the director.  The simplified hydrodynamic equation for the
director is $\partial n_i /\partial t = (K/\gamma) \nabla^2 n_i$ yields a
characteristic frequency $\omega \sim K q^2/\gamma$. The diffusion constant
$D_n = K/\gamma$ for this modes is of order $5 \times 10^{-6}$cm$^2$/s and is
much less than the diffusion constant for transverse velocity $D_v =\eta/\rho
\approx 10^{-1}$cm$^2$/sec,
where $\eta \approx 0.1$ poise is the shear viscosity and
$\rho \approx 1$ gm/cm$^3$ is the mass density.
\subsection{General Properties}
\par
The nematic example just discussed exhibits properties common to all
broken-symmetry phases.  Low-symmetry phases typically evolve from
higher-symmetry phases from which they can be distinguished by an order
parameter, which transforms under some irreducible representation of the
symmetry group $\cal G$ of the higher-symmetry phase.  If the low-symmetry
phase breaks a continuous symmetry, there are a continuum of states all
with the same free energy.  These states can be transformed into each other
under operations in $\cal G$.  They can be distinguished from each other by
some continuous (possible multi-dimensional) parameter $u$ (such as a
rotation angle $\theta$.) Spatially uniform increments in $u$ cost no
energy.  Spatially nonuniform distortions of $u$ lead to an energy cost $F
\sim K \int d^d x (\gradv u)^2$, characterized by a rigidity modulus $K$ and
implying an excitation energy $\epsilon_q = K q^2$.  Thermal
fluctuations  $\langle |u( \qv )|^2 \rangle = k_BT/K q^2$ tend to destroy
order.  Associated with each independent parameter $u$, there is an
additional low-frequency hydrodynamic mode.  We will now consider two other
examples of broken-symmetry systems.
\subsection{Smectic Liquid Crystals}
The smectic-$A$ phase can be viewed as a one-dimensional periodic stack of
fluid layers of aligned nematogens.  The ideal smectic phase is distinguished
from the nematic phase by a periodic modulation of the mass density
parallel to the director $\nv$.  Thus, an order parameter for the smectic
phase  with layers perpendicular to the $z$-axis can be defined via the Fourier
transformation of the density:
\begin{equation}
\rho ( \xv ) = \rho_0 [ 1 +  \psi e^{i q_0 [ z - u ( \xv ) ]} +
{\rm c.c.} + \cdots ] ,
\end{equation}
where $q_0 = 2 \pi / d$ and $d$ is the layer spacing.
The order parameter $\psi = |\psi | e^{- i  q_0 u }$ is the lowest-order
complex mass-density-wave amplitude.  In general, higher-order Fourier
components $\psi_n$ are needed to describe competely a periodic mass density.
However, the first component is sufficient for most smectics.  The phase
variable $u$ distinguishes different equivalent energy states: it translates
the origin of periodic modulations.
\par
The smectic free energy depends on gradients of $u$, and one might expect
it to be proportional to $(\gradv u )^2$.  This guess is, however, too
naive.  The smectic free energy must be invariant not only with respect
to spatially uniform translations but also with respect
to uniform simultaneous rotations of the
nematic director $\nv$ and the smectic layers.  This requirement leads to
the free energy
\begin{equation}
F = {1 \over 2} \int d^3 x [ B (\nabla _{||} u)^2 + D (\gradv_{\perp}u +
\delta \nv )^2 ] + F_n ,
\label{smectic1}
\end{equation}
where $||$  and $\perp$ refer, respectively,
to directions parallel and perpendicular to the $z$ axis and $F_n$ is the
Frank free energy of Eq.\ (\ref{frank}).
The combination $(\gradv_{\perp }u + \delta \nv )^2$ is invariant under
simultaneous rotations of layers and the director, whereas $(\gradv_{\perp}
u)^2$ alone is not.  The smectic-$A$ phase breaks both rotational and
translational symmetry.  One might, therefore, expect three low-energy
broken-symmetry modes associated with $u$, $\delta n_x$ and $\delta n_y$.  This
is, however, not the case.  The presence of periodic order makes it
energetically costly to rotate the director relative to the layer normal:
there is a finite energy cost $\approx (D + Kq^2 ) |\delta \nv |^2 $  as $ q
\rightarrow 0$ associated with director distortions at $\gradv u = 0$.
Layers expel director bend and twist. This phenomenon is identical to the
Higgs mechanism\cite{Higgs} for generating a massive Higgs particle in quantum
field theory.  It is also identical to the expulsion of magnetic flux in a
superconductor.
\par
If there are spatial modulations of $u$, the lowest energy
configuration for $\delta \nv $ is $\delta \nv \approx - \gradv_{\perp} u$.
Relaxing $\delta \nv$ to this value yields the Landau-Peierls free energy
for a one-dimensional solid,
\begin{equation}
F = {1 \over 2} \int d^3 x [ B (\nabla_{||} u )^2 + K_1 (\gradv_{\perp} u
)^2 ] ,
\end{equation}
leading to fluctuations, $\langle | u( \qv ) |^2 \rangle = k_BT/ (B
q_{||}^2 + K_1 q_{\perp}^4)$.  These fluctuations are so strong that the
local value of $\langle u^2 (\xv ) \rangle = \int (d^3q/(2 \pi )^3) \langle
|u ( \qv )|^2 \rangle$ diverges logarithmically with sample size.  Thus
fluctuations destroy the ideal long-range order of the smectic-$A$ phase:
$\langle \psi \rangle = |\psi | \langle e^{- i q_0 u}\rangle = |\psi | \exp
[ - q_0^2 \langle u^2 \rangle /2 ] = 0$. Even though $\psi$ is zero at any
finite temperature, the smectic phase is distinct from the nematic phase.
The rigidity $B$ is nonzero in the smectic phase and zero in the nematic
phase.
In the nematic phase, correlations in $\psi ( \xv )$ die off exponentially
with distance.  In the smectic phase, they die off algebraically\cite{Caille}:
\begin{equation}
\langle \psi( \xv ) \psi^* ( 0 ) \rangle \sim
\cases{
x_{\parallel}^{- n^2 \eta_c}, &if $\xv_\perp = 0$;\cr
|x_{\perp}|^{-2 n^2 \eta_c }, &if $x_{\parallel}=0$\,, \cr}
\label{6.4.14}
\end{equation}
where
\begin{equation}
\eta_c = {q_0^2 k_BT \over 8 \pi ( K_1 B )^{1/2}} .
\label{6.4.15}
\end{equation}
The power-law form of spatial correlations implies that the X-ray structure
factor will have a power-law rather than a delta function singularity at
$\qv = n\qv_0$:
\begin{equation}
I ( \qv ) \cases{
(q_{\parallel}-nq_0)^{-2+n^2 \eta_c}, &if $\qv_\perp = 0$;\cr
q_{\perp}^{-4+2n^2 \eta_c}, &if $q_{\parallel}=0$.\cr}
\label{6.4.16}
\end{equation}
This behavior has been observed both in single-component
smectics\cite{AlsNielsen} and in smectic phases in water-surfactant
mixtures\cite{Safinya}.
\par
The existence of a single elastic variable $u$ implies the existence of a
single additional hydrodynamic mode in a smectic compared to an isotropic
fluid.  The nature of this mode is quite interesting and contrary to
intuition based on harmonic solids.  If a smectic is translated with
uniform velocity along the $z$ axis, then the
time rate of change of the phase variable $u$ will be identical to $v_z$.
In non-equilibrium situations $\partial u/\partial t$ can differ from
$v_z$.  In other words, it is possible for molecules to flow relative to a
fixed smectic lattice.  This is the phenomenon of permeation\cite{permeation}.
There is a dissipative force opposing permeation:
\begin{equation}
{\partial u\over \partial t} - v_z = - \zeta {\delta F \over \delta u} =
\zeta (B \nabla_{||}^2 -  \nabla_{\perp}^4 ) u ,
\label{permeation}
\end{equation}
where $\zeta$ is the permeation dissipative coefficient.  The equation for
the velocity $\vv$ couples to $u$:
\begin{equation}
\rho {\partial v_i \over \partial t} = - \nabla_i p - \delta_{iz} {\delta F
\over \delta u} - \eta \nabla^2 v_i .
\label{force}
\end{equation}
These equations imply the existence of propagating shear modes with
frequencies $\omega = \pm \sqrt{B/\rho} (q_x^2 q_z^2/ q^2)$ if $q_x$ and
$q_z$ are both nonzero.  At $q_x = 0$, there is a pure permeation mode with
$\omega = - i \zeta B q_z^2$ in addition to two transverse shear modes.
\subsection{Blue Phases}
Liquid crystalline blue phases\cite{bluephases} are remarkable.  They are
periodic crystalline phases that Bragg scatter in the visible implying unit
cell dimensions of order the wavelength of light and, therefore, an enormous
number of molecules per unit cell.  Even more surprising is that there is
no discernible modulation in their mass density.  Rather there is a
periodic modulation in the nematic order parameter $Q_{ij}$ specifying the
magnitude and direction of orientational order.  Blue phases are,
nonetheless, true periodic crystals that support shear\cite{blueshear}
and form faceted surfaces\cite{facets}
when they coexist with the isotropic phase.  Order parameters for
the Blue phases can be obtained from the Fourier expansion of $Q_{ij}$:
\begin{equation}
Q_{ij} ( \xv ) = \sum_{\Gv}Q_{ij}^{\Gv}  e^{i \Gv \cdot (\xv - \uv )} ,
\end{equation}
where $\Gv $ is a reciprocal lattice vector of the cubic lattice.  The
complex tensor amplitude $Q_{ij}^{\Gv} e^{-i \Gv \cdot \uv}$ are the
blue-phase order parameters.
\par
The energy of a blue phase is invariant with respect to uniform
translations of $\uv$ and with respect to simultaneous rigid rotations of
the lattice and the anisotropic molecules.  As in a smectic, the latter
invariance introduces couplings between molecular rotation through angles
$\Omega_i$ and lattice rotations through angles $\omega_i =
\epsilon_{ijk} \partial_j u_k /2$ of the form $(\omega_i - \Omega_i )^2$
and causes molecular rotation relative to a fixed lattice to be
energetically costly.  This is the Higgs mechanism again.  As a result,
$\omega_i$ locks to $\Omega_i$ at long wavelengths, and the long wavelength
elastic energy depends only on the symmetrized strain tensor $u_{ij} =
(\partial_i u_j + \partial_j u_i )/2$ and has a form\cite{Holger1}
identical to that of any cubic crystal:
\begin{equation}
F_{\rm el} = { 1 \over 2} \int d^3 x K_{ijkl} u_{ij} u_{kl} ,
\end{equation}
where $K_{ijkl}$ is the elastic constant tensor with $3$ independent
components in a cubic crystal. The cubic anisotropy of the blue phases is
small, and $F_{\rm el}$ can be approximated by the elastic energy of an
isotropic solid with Lam\'{e} coefficients $\lambda$ and $\mu$. The shear
modulus $\mu$ is of order a Frank elastic constant divided by the square of the
lattice parameter $d \approx 2000\Angstrom$: $\mu \approx K/d^2 \approx 5
\times 10^{-7} /(2 \times 10^{-5})^2 = 10^3$ dynes/cm$^2$.  It is
$10^9$ times smaller than the shear modulus of a typical hard solid like
aluminum.
\par
The equations\cite{Holger1} governing the long wavelength hydrodynamics of a
blue phase are identical to those\cite{MPP} of any cubic crystal, hard or
soft.  The softness of the blue phases, however, leads to a mode structure
not found in hard systems.  There are three hydrodynamic variables $u_x$,
$u_y$, and $u_z$, not present in an isotropic fluid.  There are, therefore,
a total of $8$ hydrodynamical modes: two longitudinal sound modes, four
transverse sound, one heat diffusion, and one vacancy diffusion mode.
Vacancy diffusion is really mass motion relative to a fixed periodic
lattice: it is the same thing as permeation.  In hard crystals, vacancy
diffusion is a very slow activated process and can often be ignored.  In
soft crystals, especially in large unit cell structures, it cannot be.  In
addition propagating shear modes can become overdamped permeation-velocity
modes above a critical wavenumber. The displacement $\uv$ and the velocity
$\vv$ obey vector versions of the permeation and force equations, Eqs.\
(\ref{permeation}) and (\ref{force}).
The magnitude\cite{permeation,Holger1} of the
dissipative coefficient $\zeta$ is or order $d^2/[(2 \pi )^2 \gamma]$,
where $\gamma$ is the friction coefficient of a nematic liquid crystal.
These
equations predict a diffusive longitundinal permeation mode with diffusion
coefficient $D_p = \zeta \mu \approx (2 \pi)^{-2} K/\gamma \approx
10^{-7}$cm$^2$/s or about an order of magnitude smaller than the director
diffusion coefficient of the nematic and six orders of magnitude smaller than
the velocity diffusion coefficient $D_v$. Since $D_p/D_v$ is so small, it
can be ignored in the equation for shear displacements, which then obeys the
standard damped shear-wave equations: $\rho \partial^2_t \uv = - \nabla^2 (
\mu + \eta \partial_t )\uv$.  At the longest wavelengths, there is a
propagating damped shear mode with velocity
$c = \sqrt{\mu/\rho} \approx 30$cm/sec.  At
wavenumbers greater than
$q_c = 2 c/D_v = 2 \sqrt{\mu \rho}/\eta \approx 500$cm$^{-1}$ for $\eta \approx
0.1$ poise,
dissipation dominates, and the shear wave breaks up into two over-damped and
diffusive-like modes. The critical wavenumber $q_c$ is of order $10^{-4}$ times
smaller than the zone edge vector $2 \pi /d \approx 3 \times
10^6$cm$^{-1}$.  Thus, the shear mode is overdamped over most of the
Brillouin zone.  Similar behavior occurs in colloidal crystals.
\begin{figure}
\centerline{\psfig{figure=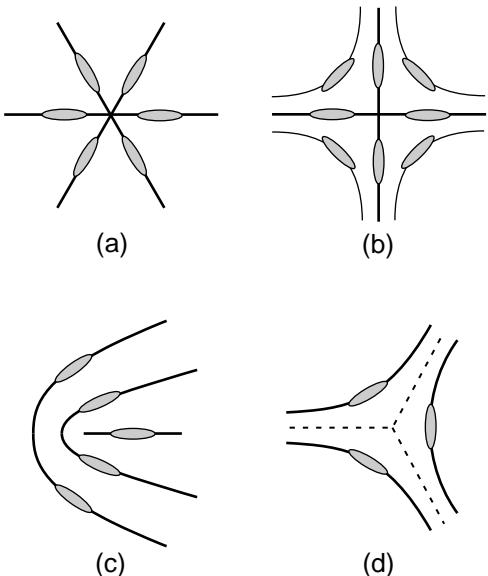}}
\caption{Director configuration for disclinations in two-dimensions with (a)
$k=1$, (b) $k=-1$, (c) $k=1/2$, and (d) $k=-1/2$.}
\label{discfig}
\end{figure}
\section{Topological Defects}
Consider a two-dimensional nematic for which the director can be written as
$\nv = (\cos \theta , \sin \theta)$.  This system can have a special kind of
defect, called a disclination, in which $\theta$ changes by $2 k \pi$, where
$k$ is a half-integer, in one circuit around a point core as shown in Fig.\
\ref{discfig}.  The order parameter
$Q_{ij}$ is continuous and well defined everywhere except at the core, even
though $\theta$ undergoes a discontinous change of $- k\pi$ along some
line originating at the core.  A disclination is one of a class of what
are called topological defects\cite{topological}.  These defects have the
property that no continuous distortion can make them disappear (i.e., return
the system to its undistorted aligned ground state).  Their properties depend
on the symmetry of the order parameter and the topological properties of the
space in which the transformation variable $\theta$ resides (the unit
circle with opposite points identified in the case of the two-dimensional
nematic).  Examples of topological defects include vortices in superfluid
helium and in superconductors, dislocations and disclinations in crystals
and liquid crystals, and hedgehogs in magnets and nematic liquid crystals.
\par
Topological defects play a determining role in a number of physically
important phenomena.  The proliferation of vortices is responsible for
the transition from the superfluid to the normal state in helium films.
Dislocations determine the strength of crystals.  The Abrikosov phase in
superconductors is a lattice of vortices.
\par
Soft condensed matter systems are ideal for the study of topological
defects.  As discussed in the preceding section, soft systems are
characterized by easy response to external forces and by characteristic
lengths much larger than a molecular scale. This makes it possible
not only to
produce defects in a controlled way via boundary conditions ond external
fields but also to produce defects with length scales that can be imaged simply
with optical techniques.  In the next two subsections, we will give some
examples of topological defects in soft systems.
\begin{figure}
\centerline{\psfig{figure=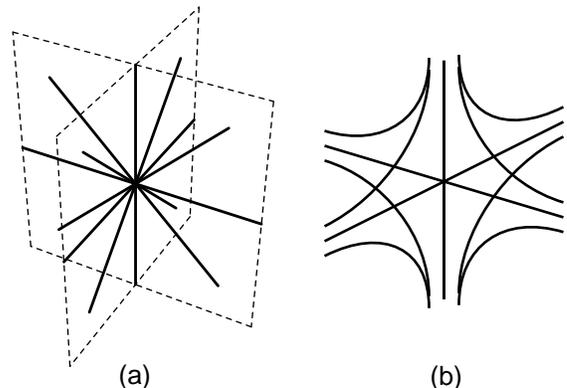}}
\caption{Director configurations for (a) a radial hedgehog with $Q= \pm 1$ and
(b) a hyperbolic hedgehog with $Q = \pm 1$.  The sign of the hedgehog in a
nematic is ambiguous because of the $\nv \rightarrow - \nv$ symmetry}
\label{hedgehogfig}
\end{figure}
\subsection{nematic liquid crystals}
\par
As we just saw, a two-dimensional nematic can have point topological
defects called disclinations.  In three-dimensional nematics, there are
strength $1/2$ disclination line defects around which the director rotates
by $(2 \pi)/2$.  In addition, there are point defects called hedgehogs (Fig.\
\ref{hedgehogfig}) characterized by a topological charge $Q$.  In the simplest
$Q =1$ hedgehog, the director points radially outward like the electric field
around a point charge. A $Q=1$ hyperbolic hedgehog is obtained from the
radial hedgehog by rotating all vectors through $\pi$ about the $z$ axis.
At low temperature, there will be a small number of
thermally activated disclination loops and hedgehogs.  As temperature
increases, the number and size of these defects will increase.  At sufficient
density, they will destroy the order of the nematic phase.  The isotropic
phase can thus be modeled as a nematic phase riddled with disclinations and
hedgehogs.  Experimental verification\cite{nematicquench} of this picture can
be obtained by suddenly quenching an isotropic phase to a temperature at which
the nematic phase is stable.  A spaghetti of disclination lines is visible
shortly after the quench.  As time goes on, the disclination density decreases
via annihilation processes.  Interestingly, the characteristic distance between
disclinations increase as a power law in time at long times.  This quenching
into the nematic phase provides a useful model for the formation of cosmic
strings\cite{nematicquench}.
\par
Another interesting example of topological defects in nematics is
provided by nematic emulsions.  If nematogens are mixed in water with a
surfactant, they will form essentially spherical emulsion
droplets\cite{Drazaic} with diameters ranging from $0.5\mu$m to
$50\mu$m and more.  Boundary conditions at the surface of the droplet depend
on the surfactant, and both normal and tangential boundary conditions can be
produced.  Normal boundary conditions\cite{topological} force the creation of a
$+1$ hedgehog at the center of the droplet.  Tangential boundary conditions
lead to surface defects called boojums\cite{boojum}.
The defect state can be modified by external
electric or magnetic fields.  Since the way a droplet scatters light depends
on its director configuration, different topological states will scatter light
differently. It is thus possible to control the amount of light scattering from
an array of droplets with modest electric fields.  This is the basis for a
display technology\cite{Drazaic}.
\begin{figure}
\centerline{\psfig{figure=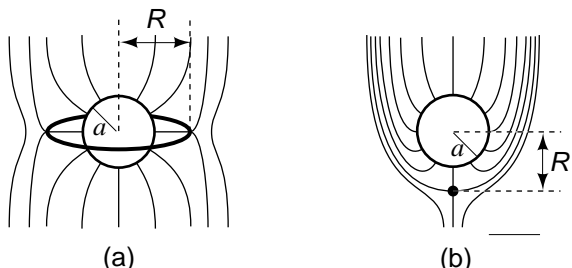}}
\caption{(a) ``Saturn Ring" configuration of a spherical particle with normal
boundary conditions in a nematic.  A disclination ring screens the radial
hedgehog created by the particle producing parallel alignment at long
distances. (b) Dipole configuration in which the additional particle creates
a hyperbolic hedgehog.}
\label{emulsion}
\end{figure}
\par
If small water droplets favoring director alignment normal to their
surfaces are dispersed in a nematic, they will nucleate a
hedgehog\cite{Poulin}.  If fields or boundary conditions force average
alignment of the nematic along some direction, the total hedgehog number must
be zero.  Thus, each water droplet must create a compensating companion defect
such as depicted in Fig.\ \ref{emulsion}.
If the diameter of the water droplet is of order a micron or larger, the
compensating defect is a hyperbolic hedgehog as shown in Fig.\
\ref{emulsion}b.  This defect leads to a short-range repulsion between
droplets that prevents coalescence.  The droplet-defect pair form a dipole
configuration that disrupts the far-field director and leads to an
attractive interaction between aligned dipoles.  The result is that water
droplets in a nematic form chains just like magnetic dipoles with
short-range repulsion.  Smaller droplets may nucleate a dsiclination ring
about their equator\cite{terent}
in the ``Saturn ring" configuration shown in Fig.\
\ref{emulsion}a.
\subsection{Smectic Liquid Crystals}
\par
Dislocations in smectics are topological defects in which $u$ undergoes
a change of $kd$, where $k$ is an integer  and $d$ is the smectic layer
spacing, in one circuit around a linear core. Dislocations form when a smectic
is forced into geometric environments that are incompatible with its layered
structure.  For example there must be dislocations in a smectic wedged between
two nonparallel surfaces if boundary conditions force the director to be
normal to those surfaces.
\begin{figure}
\centerline{\psfig{figure=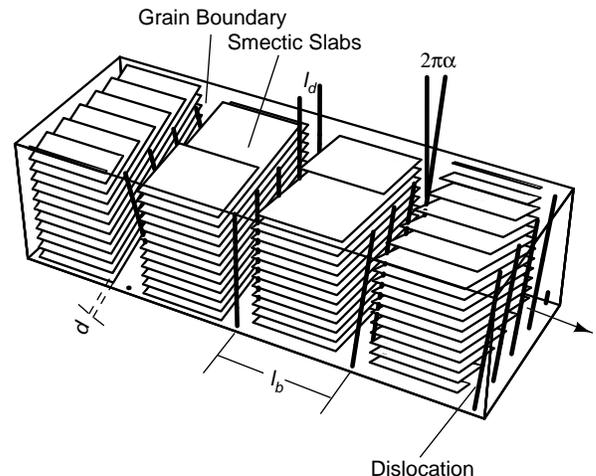}}
\caption{Schematic representation of the TGB phase showing smectic slabs
separated by grain boundaries composed of periodically spaced screw
dislocations.  The change in angle of the smectic layer normal across a
grain boundary is $2 \pi \alpha = 2 \sin^{-1} d/2 l_d$ where $d$ is the
layer spacing and $l_d$ is the separation between dislocations.  If
$\alpha$ is a rational number $P/Q$ for relatively prime integers $P$ and
$Q$, the structure has a $Q$-fold screw axis and quasicrystalline symmetry
if $Q=5$ or $Q>6$.  If $\alpha$ is irrational, the structure is
rotationally incommensurate.}
\label{TGBfig}
\end{figure}
\par
A particularly beautiful example of topological defects in smectic
liquid crystals is the twist-grain-boundary or TGB phase\cite{TGB} shown in
Fig.\ \ref{TGBfig}.  Chiral molecules prefer twisted to parallel structures.
If chirality is added to a nematic liquid crystal, a twisted nematic or
cholesteric phase results in which $\nv = ( \cos k_0 z , \sin k_0 z, 0
)$ rotates in a helical fashion about a pitch axis.  This twisting
structure is incompatible with the periodic layering of the smectic
phase.  Thus the smectic either expels twist altogether, or it permits
twist via the formation of a regular lattice of grain boundaries, each
composed of a periodic array of screw dislocations, across which the
smectic layers undergo a discrete rotations.  The smectic phase is
analogous the superconducting phase, and the nematic--to--smectic-$A$
transition is analogous to the the normal-to-superconduction
transition\cite{deGennes}. Twist is the analog of the
$B$ field in a superconductor.  The TGB phase is the analog of the
Abrikosov vortex lattice phase and the smectic phase with expelled twist
is the analog of the Meissner phase in superconductors.  TGB phase have
been found in a large number of compounds\cite{TGBphases}.  There are even
quasicrystalline\cite{Bordeaux} TGB phases with $16$- to $20$-fold screw axes.
\section{Membranes}
Membranes\cite{membranes} are two-dimensional flexible surfaces.
Aliphatic molecules, such
as detergents and biologically important phopholipids, are molecules with
a polar hydrophilic (water-loving) head and a hydrocarbon hydrophilic
(water-fearing) flexible tail.  When these molecules are disolved in
water, they spontaneously form structures that protect the hydrocarbon
tail from contact with water.  At sufficient concentration, they form
large, bilayer membranes that can assume a variety of shapes including
nearly flat surfaces, closed vesicles with the topology of a sphere, and
more complex multi-connected surfaces similar to the Fermi surface of
copper.  These membranes are both flexible and fluid: their shape changes
easily, and their constituent molecules diffuse freely within the confines of
the surfaces they define.  They are perfect examples of soft materials.
\par
In its lowest energy state, the membrane defines a flat surface that
breaks the translational and rotational symmetry of space.  If the fluid
membrane is fee and not under tension, the energy of deformed sates is
determined by the Helfrich-Canham\cite{HelCan} free energy:
\begin{equation}
F_{\rm mem} = {1 \over 2 } \kappa \int dS \left( {1 \over R_1} + {1 \over
R_2} \right)^2 \approx {1 \over 2} \kappa \int dS (\nabla^2 h )^2 ,
\end{equation}
where $R_1$ and $R_2$ are the local radii of curvature, $R_1^{-1} +
R_2^{-2}$ is the local mean curvature, and $\kappa$ is a bending rigidity
with units of energy.  The second form of this energy is the harmonic
expansion about the flat surface with points on the membrane parametrized
by their height $h(x,y)$ above the $xy$ plane.  The rigidity $\kappa$ has
values that range from one to several tens of $k_B T$.  $F_{\rm mem}$ is
invariant with respect to both uniform translations and rotations
The layer normal is $\Nv \approx ( - \partial_x h , -\partial_y h , 1 )$
for small $\nabla h$.  Thus $\gradv h$ describes a rotation of $\Nv$,
which cannot change the membrane energy. There can be no $(\nabla h)^2$
contribution of $F_{\rm mem}$, and the leading contribution is $(\nabla^2
h )^2$.  A membrane has all of the modes of a two-dimensional
fluid (two-dimensional sound and heat and momentum diffusion) plus a height
mode with a long-wavelength energy $\epsilon_q = \kappa q^4$
proportional to $q^4$ (rather than $\sim q^2$) because both translational
and rotational symmetry are broken.  The latter mode dissipates energy into the
shear modes of the surrounding fluid.  It is overdamped with a
frequency $\omega_q$ that must decrease with increasing fluid viscosity
$\eta$.  Dimensional analysis then implies that $\omega_q \sim (\kappa /
\eta)q^3 \approx 10^4$Hz at $\lambda = 2 \pi / q \approx 2000 \Angstrom$.
Thermal height fluctuations $\langle |{\delta h ( \qv ) }|^2 \rangle = k_B T
/\kappa q^4$ diverge strongly at small $q$ and lead to a decorrelation of
layer normals.  A persistence length $L_p$ beyond which orientational memory is
lost can be defined at the length at which the layer-normal correlation
function,
\begin{eqnarray}
\langle [\Nv ( \xv ) - \Nv ( 0 )] \rangle & = &
{k_b T \over \kappa} \int {d^2 q
\over ( 2 \pi )^2} {1 - e^{i \qv \cdot \xv }\over q^4} \nonumber \\
& = & {k_B T \over 2
\pi \kappa} \ln |\xv | / \xi ,
\end{eqnarray}
becomes of order unity\cite{taupin}.
This yields $L_p = \xi e^{2 \pi \kappa/k_B T}$.  At length scales beyond
$L_p$ a free membrane will no longer be flat; it will assume complicated
crumpled configurations.
\par
\begin{figure}
\centerline{\psfig{figure=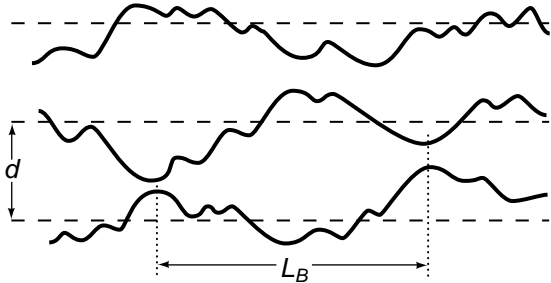}}
\caption{Representation of a lattice of fluctuating membranes.  The average
separation between membranes is $d$.  The average distance in the $xy$
plane between collisions is $L_B$.  Between collisions, each membrane is
free.  The distance $d$ and $L_B$ are related via $d^2 = (k_B T/\kappa)
L_B^2$.}
\label{collisions}
\end{figure}

Fluid membranes are the basic constituents of a number of
thermodynamically stable phases of which the simplest is the lyotropic
lamellar phase formed by a periodic stack of membranes separated by water.
This phase is a
smectic liquid crystalline phase with a layer spacing $d$ that can be
varied from $50 \Angstrom$ to several microns.  There is a negative
interfacial energy per unit area $\sigma$
that favors large membrane area.  This is opposed by
interactions between membranes.  When charges are fully screened,
membranes interact only when they are in close contact.  Height
fluctuations of a membrane confined between two others will be smaller
than those of a free membrane.  The confined membrane has a lower entropy
than a free membrane. As a result, there will be an entropic repulsion
between membranes.  The balance between the attractive interfacial force
and the entropic repulsive entropic force determines the equilibrium layer
spacing and the elastic compression modules $B$.  Helfrich\cite{helfrich2}
provided an estimate of the entropy reduction as follows.  A fluctuating
membrane will collide with its neighbors as depicted in Fig.\
\ref{collisions}.  The mean-square height fluctuation will be of order
$d^2$ at a length scale of order $L_B$, the mean distance between
collisions in the $xy$ plane:
\begin{equation}
d^2 = \langle h^2 ( \xv ) \rangle \approx {k_B T \over \kappa} L_B^2
{}.
\end{equation}
Each collision leads to an entropy reduction of order $k_B$.  There are
$N = V/(L_B^2 d )$ collisions in a volume $d$.  Thus the entropy
reduction per unit volume is $\Delta s = - k_B / L_B^2 d$.  The gain
in free energy per unit volume arising from increased surface area is $-
\sigma /d$.  Thus the free energy density of a stack of membranes is
\begin{equation}
f = {(k_B T)^2\over \kappa d^3} - {\sigma \over d} .
\end{equation}
Minimization with respect to $d$ yields $ d = (\sigma \kappa / 3 k_B^2 T^2
)^{1/3}$ and a compression modulus $B = d^2 \partial^2 f /\partial d^2 = 6
k_B^2 T^2 /\kappa d^3$.  X-ray experiments have verified this
relation\cite{Safinya}.  An almost identical entropic analysis applies to
the striped phase of atoms adsorbed onto a graphite
substrate\cite{graphite}.
\par
Membranes can develop varying degrees of internal order that modify their
properties\cite{membranes}.
For example, their constituent molecules can tilt relative
to their normals.  This breaks rotational symmetry in the plane of the
membrane and leads to a vector in-plane orientational order parameter.
Or, polymerization can produce a two-dimensional solid with a harmonic
shear and bulk modulus but with height fluctuations out of the plane of
the membrane.  As in a fluid, the latter
fluctuations are very soft with
energy $\epsilon_q = \kappa q^4$.  Coupling between the soft height mode
and in-plane shear leads to a remarkable renormalization\cite{polmem} of the
elastic so that the renormalized long-wave-length bending modules diverges
as $q^{- \eta_h}$ and the shear and bulk moduli vasish as $q^{\eta_u}$
with $\eta_u = 2 ( 1 - \eta_h )$.  Non-linearities actually destroy
harmonic elasticity!
\section{Prospects for the Future}
\par
Soft condensed matter physics is a vast and vibrant field.  It will
continue to be a growth area for the foreseeable future enriching both
physics and the many sciences such as chemistry, chemical engineering,
materials science, and biology that it overlaps.  Listed below are some
(but certainly not all) areas where one can expect to see exciting progress in
the next few years.
\subsection{New Structures}
\par
The ease with which soft condensed matter can deform is responsible for
such remarkable phases as the TGB phase.  There are surely others to be
discovered.  For example, disc-like (rather than rod-like) molecules or
semiflexible polymers tend to form columnar structures in which there is
hexagonal crystalline order in two dimensions and fluid-like structure in
the third.  Chirality in these systems should produce a variety of
``braided" and TGB-like structures\cite{kamnel}.
A good candidates system to see these
phase is aligned DNA.  Another structure that may exist is a TGB-blue
phase in which smectic layering coexists with a three-dimensional twist
structure.  The ability of synthetic chemists to engineer molecules with
exotic shapes plays an important role in this arena.
\subsection{Measurement and Control at the Micron Scale
and Lower}
\par
A variety of new or improved experimental techniques. including laser and
magnetic tweezers and fluorescence and near-field microscopy make it
possible both to visualize and to control processes at the micron scale
and lower.  For example, laser tweezers can be used to confine colloidal
particles to specified regions, to move them about, and to measure
piconewton forces.  One can expect to see an explosion of new experimental
data on a variety of systems.  Examples of experiments that have already
been done include the measurement of extension versus force on
DNA\cite{DNA}, the effect of depletion forces on diffusion in controlled
geometries\cite{Arjun}, and the laser induction of pearling
instabilities in bilayer cylindrical vesicles\cite{pearl}.  More will
follow.
\par
This new control will also lead to new materials.  In the near future, we
should see designer two- and three-dimensional colloids engineered through
clever use of surface templates, depletion forces, laser tweezers, and
related techniques.  Interesting new materials would be optical band gap
materials in the form of a regular $3$D lattice of low and high dielectric
constant spheres or a $3$D crystal of two different size nematic emulsion
droplets.
\par
Nanoscale phenomena is a hot subject in hard (electronic) as well as soft
condensed matter physics.  Soft condensed matter will be used to create
templates for the creation of metallic nanostructures.
\subsection{Biology}
\par
One of the most exciting areas of soft condensed matter physics is its
interface biology.  The fundamental building blocks, the plasma membrane, the
cytoskeleton, microtubules, DNA and actin molecules, etc.,
are soft materials.  They have mechanical properties that
are well described by the language presented in
this article:  They are polymers or surfaces with differing rigidities;
they are subject to depletion forces and viscous forces when they move,
etc.  Soft condensed matter physics will have an increasing impact on
biology and conversely biology,
by providing examples of how nature creates and uses
structures, will provide paradigms for new soft materials.
\par
I am pleased to dedicate this article to Eli Burstein on the occasion of
his 80th birthday.  His love of and commitment to all physics, and
condensed matter physics in particular, has been an inspiration to us all.
\par
The author is grateful to Arjun Yodh and David Weitz for useful
conversations.  This work was supported in part by grants from NSF under
grant No. DRM94-23114 and DMR91-22645.


\begin{references}
\bibitem{glass}  There are of course many random-close-packed
configurations that are ultimately responsible for the ``ground state
entropy" of a galss.  Different configurations are, however, inaccessible
to hard spheres at true close pakcing, and we may take to entropy of a
single configuration to be zero.
\end{references}
\end{document}